\documentclass[12pt,epsf,letterpaper]{article}%
\usepackage{amsmath}
\usepackage{geometry}

\usepackage{graphicx}
\begin{document}
\title{{\bf Chern--Simons--Yang--Mills system in presence of Gribov horizon with fundamental Higgs matter}}
\author{$^{(1)}$Arturo J. Gomez, $^{(2)}$Sebastian Gonzalez, $^{(3)}$Silvio Paolo Sorella $$\\\textit{$^{(1)}$Departamento de Ciencias, Facultad de Artes Liberales y}\\\textit{Facultad de Ingenier\'{\i}a y Ciencias, Universidad Adolfo
Ib\'{a}\~{n}ez, Vi\~{n}a del Mar.} \\\textit{$^{(2)}$Scientific Division, Patio Baron Foundation, Valparaiso} \\\textit{$^{(3)}$ Instituto de F\'{\i}sica, Universidade do Estado do Rio de Janeiro} \\\textit{\ Rua sao Francisco Xavier 524, Maracana, Rio de Janeiro.}}
\maketitle

\begin{abstract}
In this work we study the behaviour of Yang--Mills--Chern--Simons theory coupled to a Higgs field in the fundamental representation by taking into account the effects of the presence of the Gribov horizon. By analyzing the infrared structure of the gauge field propagator, both confined and de-confined regions can be detected. The confined region  corresponds to the appearance of complex poles in the propagators, while the de-confined one to the presence of real poles. One can move from one region to another by changing the parameters of the theory.  
\end{abstract}

\section{Introduction} 

It is widely known that the issue of the Gribov copies \cite{Gribov:1977wm} in
non-abelian gauge theories is deeply related to the problem of color
confinement. In  recent years the Gribov--Zwanziger (GZ) approach \cite{Zwanziger:1989mf,Vandersickel:2012tz} has become a promising framework in order to describe several
features of the infrared regime of Yang--Mills  theories. For instance, the 
inclusion of dimension two condensates  \cite{Dudal:2007cw,Dudal:2008sp}   provides a
refinement of the Gribov-Zwanziger framework  allowing to describe
the infrared behaviour of the gluon propagator in very good agreement  with the
most recent lattice data \cite{Cucchieri:2007rg,Cucchieri:2007md}. Within this
approach, it is possible to investigate the spectrum of 
Yang--Mills theories by constructing local gauge invariant composite operators suitable to estimate  
the masses of the lightest glueball states \cite{Dudal:2010cd,Dudal:2013wja}. Also, the so-called Refined Gribov
Zwanziger action \cite{Dudal:2007cw,Dudal:2008sp} has been used in the computation of the Casimir energy in
the context of the MIT--bag model \cite{Canfora:2013zna}, providing the correct sign for the Casimir force.
A very important problem related to the
non--perturbative behavior of Yang--Mills theory is the transition between
confining and non--confining phases in presence of Higgs fields, see \cite{Polyakov:1976fu,Cornwall:1998pt,Baulieu:2001vw}. For
instance, in three dimensions, Polyakov's seminal work \cite{Polyakov:1976fu} shows that in the
Georgi--Glashow model monopole configurations in the Euclidean space give a
dominant contribution in the functional integral, providing a successfull
mechanism for  confinement, in agreement with the dual superconductivity picture.

\bigskip

\noindent Gauge--Higgs systems for the gauge group $SU(2)\ $  within  the context of the Gribov
problem have been addresses in   \cite{Capri:2012cr,Capri:2012ah,Capri:2013oja}. A very 
interesting aspect to be mentioned is that concerning  the physical consequences in the phase
structure of the theory due to the choice of the adjoint or of the fundamental
representation for the Higgs field. In particular, in the case of the Georgi--Glashow
model, {\it i.e.} of the three-dimensional Yang-Mills theory with Higgs fields in the adjoint representation, 
the third component of the gauge field, $A_{\mu }^{3}$, is always
confined for all values of the gauge coupling $g$ and of the vev $v$ of
the Higgs field. Within the Gribov approach, this feature turns out to be encoded in the corresponding gluon propagator which is 
of the Gribov type, {\it i.e.} it displays two unphysical complex conjugate poles. 
Moreover, the off-diagonal gauge propagator  $\left\langle A_{\mu }^{\alpha }(q)A_{\nu }^{\beta
}(q)\right\rangle \ $, with $\alpha,\beta=1,2$, decomposes into the sum of two Yukawa propagators with
real and positive masses. Though, only the heaviest mass component of this
decomposition has a positive residue and can be regarded as a physical mode which is, however, 
decoupled form the infrared dynamics due to its large mass. In the case of Higgs fields in the 
fundamental representation, the gauge group $SU(2)$ is completely broken. At
weak coupling,  all propagators decompose into a sum of two Yukawa
propagators with positive masses. One of the components is unphysical due to
a negative residue. However, the component with the largest mass is 
a physical mode. Therefore, at weak coupling all gauge modes display a massive
physical component. This is what can be called a Higgs phase. In the strong
coupling, the propagator of all gauge modes are of the Gribov type,
exhibiting complex conjugate poles. This is the confining phase. Therefore,
when the Higgs field is in the fundamental representation, we have a weak
coupling Higgs phase and a strong coupling confining phase. These results are in nice agreement with the behavior reported by
lattice simulations  \cite{Fradkin:1978dv,Nadkarni:1989na}.

\bigskip

\noindent A similar behaviour in the infrared region is observed when considering a
Yang--Mills field coupled to a Chern--Simons topological term \cite{Deser:1981wh} in the Landau gauge \cite{Canfora:2013zza}. 
Unlike  pure three--dimensional Yang--Mills theory, where the effect of the Gribov
horizon confines all  degrees of freedom of the theory, 
the addition of the Chern-Simons  topological term allows for  the possibility 
of having physical poles in the resulting propagator gauge for certain values of the coupling
constant $g$ and of the topological mass $M$, making possible to identify regions of
confinement and de-confinement in the parameter space of the model.

\bigskip

\noindent In this work we pursue our previous investigation \cite{Canfora:2013zza} by studying Yang--Mills--Chern--Simons--Higgs systems when the 
the presence of the Gribov horizon is taken into account. 
\bigskip 

\noindent The paper is organized as follows. In Sect.2 we study the Yang-Mills-Chern-Simons-Higgs system with scalar fields in the fundamental 
representation. Sect.3 collects our conclusion.


\section{Chern-Simons-Yang-Mills-Higgs model in the fundamental representation
of SU(2)}

Let us start this section by reminding briefly Gribov's procedure \cite{Gribov:1977wm} in order to take into account the presence of gauge copies in the functional Euclidean integral. It  amounts to restrict the domain of integration in the path integral to the so called
 Gribov region $\Omega$, defined as the set all all field configurations fulfilling the Landau gauge, $\partial_{\mu}A_{\mu}^{a}=0$, and for which the Faddeev-Popov operator 
$ {\mathcal{M}}^{ab}=-\partial^2\delta^{ab}-gf^{abc}A_{\mu}^{ab}$ is strictly positive, namely 
\begin{equation} 
 \Omega =\left\{A_{\mu}^{a}; \partial_{\mu}A_{\mu}^{a}=0; \mathcal{M}=-\partial^2\delta^{ab}-gf^{abc}A_{\mu}^{ab}>0\right\} \;.  \label{omega} 
\end{equation} 
The regione $\Omega$ is known to be bounded in all directions in field space and to be convex. The boundary $\partial \Omega$ of the Gribov region is known as the Gribov horizon, where the first vanishing eigenvalue of the operator  $ {\mathcal{M}}^{ab}$ shows up. Furthermore, it has been shown that every gauge orbit crosses at least once the region $\Omega$, thus giving a well defined support to Gribov's original proposal to cut off  the functional integral at the Gribov horizon. 

\bigskip 

\noindent In order to implement the restriction to the region $\Omega$, Gribov imposed  the no-pole condition  \cite{Gribov:1977wm} on the Faddeev-Popov ghost propagator, which is nothing but the inverse of the operator ${\mathcal{M}}^{ab}$. More precisely, following \cite{Gribov:1977wm}, one parametrizes the ghost propagator $\mathcal{G}(q,A)$ as 
\begin{equation}
\mathcal{G}^{ab}(q,A)= \langle q| \left( \frac{1}{{\mathcal{M}}} \right)^{ab} |q\rangle = \frac{\delta^{ab}}{q^2}\frac{1}{1-\sigma(q,A)} \;, \label{ghost}
\end{equation} 
and one imposes the  condition 
\begin{equation}
\sigma(q,A) < 1 \;, \label{npole}
\end{equation}
which ensures that the inverse of the Faddeev-Popov operator ${\mathcal{M}}^{ab}$ is always positive, so that one always remains inside  the Gribov region $\Omega$, {\it i.e.} the Gribov horizon  $\partial \Omega$  is never crossed. As the form factor $\sigma(q,A)$ turns out to be a decreasing function   \cite{Gribov:1977wm} of the momentum $q$, it is sufficient to require 
\begin{equation}
\sigma(0,A) <  1 \;, \label{no-pole}
\end{equation}
which is known as the Gribov no-pole condition \cite{Gribov:1977wm}. According to the no-pole prescription \eqref{no-pole}, the Faddeev--Popov quantization formula gets modified as \cite{Gribov:1977wm}:
\begin{eqnarray}
d\mu_{FP}&=&\mathcal{D}A\;\delta(\partial A)\;det(\mathcal{M}^{ab})\;e^{-S_{YM}}\\ \nonumber
&\longrightarrow& \mathcal{D}A\;\delta(\partial A)\;det(\mathcal{M}^{ab})\;\theta(1-\sigma(0,A))\;e^{-S_{YM}}
\end{eqnarray}
where $S_{YM}$ is the Yang--Mills action 
\begin{equation}
S_{YM}=\frac{1}{4}\int d^4x\; F^a_{\mu\nu} F^a_{\mu\nu}  \;, \label{ymact} 
\end{equation}
and $\theta(x)$ stand for the step function. Making use of the integral representation
\begin{equation}
\theta(x)=\int_{-i\infty+\epsilon}^{+i\infty+\epsilon}\frac{d\beta}{2\pi i\beta}e^{-\beta x},
\end{equation}
it turns out that the ghost form factor $\sigma(0,A)$ can be brought into the exponential of the Yang--Mills measure $d\mu_{FP}, i. e.$ 
\begin{equation}
e^{-S_{YM}}\longrightarrow e^{-S_{YM}}e^{\beta\sigma(0,A)}  \;. 
\end{equation}
Moreover, making use of the saddle point in order to evaluate the integral over $\beta$ \cite{Gribov:1977wm}, for the partition function $\mathcal{Z}$, one writes
\begin{equation}
\mathcal{Z}=\int\mathcal{D}A \; \delta(\partial A) \; det(\mathcal{M}^{ab})\; e^{-S_{YM}}e^{\beta^*(1-\sigma(0,A))},
\end{equation}
where, to the first order,  $\beta^*$ is determined by the gap equation \cite{Gribov:1977wm}
\begin{equation}
1=\frac{3Ng^2}{4}\int\frac{d^4q}{(2\pi)^4}\frac{1}{q^4+\gamma^4 }   \;,   \qquad   \gamma^4 = \frac{g^2N}{2(N^2-1)}\beta^*  \;. 
\end{equation}
Having given a short account of the main steps of Gribov's construction, let us focus on the action of the topologically massive Yang-Mills \cite{Deser:1981wh} coupled to a Higgs field in the fundamental representation, namely
\begin{eqnarray}
S &=&S_{CS}+S_{FP}+S_{\phi } \nonumber \\
&=&-iM\int d^{3}x\;\epsilon _{\mu \rho \nu }\left( \frac{1}{2}A_{\mu
}^{a}\partial _{\rho }A_{\nu }^{a}+\frac{1}{3!}gf^{abc}A_{\mu }^{a}A_{\rho
}^{b}A_{\nu }^{c}\right) +\frac{1}{4}\int d^{3}x\;F_{\mu \nu }^{a}F_{\mu \nu
}^{a} \nonumber \\ 
&+& \int d^{3}x\left( b^{a}\partial _{\mu }A_{\mu }^{a}+\bar{c}%
^{a}\partial _{\mu }D_{\mu }^{ab}c^{b}\right)  
+\int d^{3}x\left( D_{\mu }^{ij}\Phi ^{j}D_{\mu }^{ik}\Phi ^{k}+\left(
\Phi ^{\dagger }\Phi -\nu ^{2}\right) ^{2}\right) \;,   \label{tymhfd}
\end{eqnarray}
where $b^a$ stands for the Lagrange multiplier implementing the Landau gauge, $\partial_\mu A_\mu^a=0$, and $({\bar c}^a,c^a)$ are the Faddeev-Popov ghosts.  In the fundamental representation, the covariant derivative is defined by
\begin{equation}
D_{\mu }^{ij}\Phi ^{j}=\partial _{\mu }\Phi ^{i}-ig\frac{\left( \tau
^{a}\right) ^{ij}}{2}A_{\mu }^{a}\Phi ^{j}
\end{equation}
where $i,j=1,2.$  refer to the fundamental representation of $SU(2)$ and $\tau ^{a}$
are the Pauli matrices. When $\Phi $ acquires a vacuum expectation value, we
can use the freedom of the $SU(2)$ rotations to write this expectation value in
the form
\begin{equation}
\left\langle \Phi \right\rangle =\frac{1}{\sqrt{2}}\left( 
\begin{array}{c}
0 \\ 
\nu%
\end{array}%
\right)
\end{equation}
As a consequence,  all  components of the gauge field acquire the same mass $m^{2}=%
\frac{g^{2}\nu ^{2}}{4}$. 

\subsection{Infrared behaviour gauge field propagator}
After the spontaenous symmetry breaking, the quadratic part of the action is

\begin{equation}
S_{quad}=\int d^{3}x\left( \frac{1}{4}\left( \partial _{\mu }A_{\nu
}^{a}-\partial _{\nu }A_{\mu }^{a}\right) ^{2}-i\frac{M}{2}\epsilon _{\mu
\rho \nu }A_{\mu }^{a}\partial _{\rho }A_{\nu }^{a}+b^{a}\partial _{\mu
}A_{\mu }^{a}+\frac{g^{2}\nu ^{2}}{8}A_{\mu }^{a}A_{\mu }^{a}\right)
\end{equation}
The generalization of the implementation of the restriction to the Gribov region $\Omega$ to the action  \eqref{tymhfd} can be done by following the procedure outlined at the beginning of this section. Taking into account the effects of the Gribov horizon, for the 
the gauge propagator one obtains
\begin{equation}
\left\langle A_{\mu }^{a}(q)A_{\nu }^{b}(-q)\right\rangle =\delta ^{ab}\frac{%
q^{2}(\gamma ^{4}+q^{4})+g^{2}\nu ^{2}q^{4}}{M^{2}q^{6}+(\gamma
^{4}+q^{4})^{2}+2g^{2}\nu ^{2}q^{2}\left( \gamma ^{4}+q^{4}\right) }\left(
\delta _{\mu \nu }-\frac{q_{\mu }q_{\nu }}{q^{2}}+\frac{M\epsilon _{\mu \rho \nu }q_{\rho } \;q^{2}}{\gamma
^{4}+q^{4}+g^{2}\nu ^{2}q^{2}}\right)
\label{Prop}
\end{equation}
It is easy to check that in the cases when $M=0,$ $\gamma =0$ or $\nu =0,$ one recovers  the propagators studied previously in \cite{Canfora:2013zza,Capri:2012cr}.
In  particular, in \cite{Canfora:2013zza},  we have shown that the Chern-Simons term
doesn't contribute to the Gribov gap equation, due to its topological nature. As a consequence, in the present case, for the 
gap equation determining the value of the Gribov parameter $\gamma$, we have  \cite{Capri:2012cr}
\begin{equation}
\frac{4}{3}g^{2}\int \frac{d^{3}q}{\left( 2\pi \right) ^{3}}\frac{1}{q^{4}+%
\frac{g^{2}\nu ^{2}}{4}q^{2}+\gamma ^{4}}=1  \label{gapeq}
\end{equation}
which gives%
\begin{equation}
\gamma ^{4}=\frac{1}{4}\left( \frac{g^{2}\nu ^{2}}{4}-\frac{g^{4}}{9\pi ^{2}}%
\right) ^{2}  \label{gapcond}    \;. 
\end{equation}
The gap condition \eqref{gapeq}  determines the Gribov parameter $\gamma $ in terms of the 
coupling constant $g$ and of the vacuum expectation value of the Higgs field $\nu $\\\\Therefore, taking into account the condition  \eqref{gapcond}  and looking at the propagator $\left( \ref%
{Prop}\right) $,  one sees that its analytic structure depends on the three
parameters $(M,g,\nu ).\ $ Nevertheless, it  turns out to be useful to define dimensionless generalized
variables by absorbing the Chern Simons mass in the quantities $\left(
q_{\mu },\gamma ,g^{2},\nu ^{2}\right) \ $ of the propagator, {\it i.e.} by introducing the rescaled quantities 
$\left(k_{\mu },{\tilde \gamma} , {\tilde g}^{2}, {\tilde \nu}^{2}\right) \ $

\begin{eqnarray}
q_{\mu } &=&Mk_{\mu } \\
\gamma &=&M\tilde{\gamma}  \notag \\
g^{2} &=&M\tilde{g}^{2}  \notag \\
\nu ^{2} &=&M\tilde{\nu}^{2}  \label{Redef}  \;, 
\end{eqnarray}
so that for the gauge propagator we have
\begin{eqnarray}
\mathcal{G}_{\mu \nu }^{ab}(k)&=&\left\langle A_{\mu }^{a}(k)A_{\nu
}^{b}(-k)\right\rangle =\frac{\delta ^{ab}}{M^{2}}\frac{k^{2}(\tilde{\gamma}%
^{4}+k^{4})+\tilde{g}^{2}\tilde{v}^{2}k^{4}}{k^{6}+(\tilde{\gamma}%
^{4}+k^{4})^{2}+2\tilde{g}^{2}\tilde{v}^{2}k^{2}\left( \tilde{\gamma}%
^{4}+k^{4}\right) }  \nonumber \\ 
&\times&\left( \delta _{\mu \nu }-\frac{k_{\mu }k_{\nu }}{k^{2}}+%
\frac{k^{2}}{\tilde{\gamma}^{4}+k^{4}+\tilde{g}^{2}\tilde{v}^{2}k^{2}}%
M\epsilon _{\mu \rho \nu }k_{\rho }\right)  \label{gprop}  \;, 
\end{eqnarray}
in which the parameter $M$ appears as a global factor. Of course,  expression \eqref{gprop} is valid only for $M\neq 0$.   \\\\As we have already discussed, the gap equation \eqref{gapcond} yields the parameter $\tilde{\gamma}$ as a
function of the parameters $(\tilde{g},\tilde{\nu})$. This feature allows us
to describe the analytic structure of the propagator \eqref{gprop}
 in terms of two dimensionless parameters $(\tilde{g},\tilde{\nu})$%
,\ by analizing the poles of expression \eqref{gprop}. \\\\ To discuss the  properties of the poles of \eqref{gprop},  
we first rewrite expression it as 
\begin{equation}
\mathcal{G}_{\mu \nu }^{ab}(k)=\left. \mathcal{G}_{\mu \nu }^{ab}(k)\right\vert _{par}+\left. \mathcal{G}_{\mu \nu }^{ab}(k)\right\vert _{par-viol}\end{equation}
with
\begin{eqnarray}
\left. \mathcal{G}_{\mu \nu }^{ab}(k)\right\vert _{par} &=&\frac{\delta ^{ab}%
}{M^{2}}\frac{k^{2}(\tilde{\gamma}^{4}+k^{4})+\tilde{g}^{2}\tilde{v}^{2}k^{4}%
}{k^{6}+(\tilde{\gamma}^{4}+k^{4})^{2}+2\tilde{g}^{2}\tilde{v}%
^{2}k^{2}\left( \tilde{\gamma}^{4}+k^{4}\right) }\left( \delta _{\mu \nu }-%
\frac{k_{\mu }k_{\nu }}{k^{2}}\right) \;, \label{pc}\\
\left. \mathcal{G}_{\mu \nu }^{ab}(k)\right\vert _{par-viol} &=&\frac{\delta
^{ab}}{M^{2}}\frac{k^{4}}{k^{6}+(\tilde{\gamma}^{4}+k^{4})^{2}+2\tilde{g}^{2}%
\tilde{v}^{2}k^{2}\left( \tilde{\gamma}^{4}+k^{4}\right) }M\epsilon _{\mu
\rho \nu }k_{\rho }  \;,\label{pv}
\end{eqnarray}
where $\left. \mathcal{G}_{\mu \nu }^{ab}(k)\right\vert _{par}$ and $\left. \mathcal{G}_{\mu \nu }^{ab}(k)\right\vert _{par-viol}$ stand, respectively, for the parity conserving and parity violating part of the gauge propagator \eqref{gprop}. Further, we decompose the corresponding denominators in partial fractions, obtaining  their pole structure 
\begin{eqnarray}
\left. \mathcal{G}_{\mu \nu }^{ab}(k)\right\vert _{par-viol} &=&\delta
^{ab}\left( \frac{\mathcal{R}_{1}}{k^{2}+m_{1}^{2}}+\frac{\mathcal{R}_{2}}{%
k^{2}+m_{2}^{2}}+\frac{\mathcal{R}_{3}}{k^{2}+m_{3}^{2}}+\frac{\mathcal{R}%
_{4}}{k^{2}+m_{4}^{2}}\right) M\epsilon _{\mu \rho \nu }k_{\rho } \label{pole1}\\
\left. \mathcal{G}_{\mu \nu }^{ab}(k)\right\vert _{par} &=&\delta
^{ab}\left( \frac{\mathcal{F}_{1}}{k^{2}+m_{1}^{2}}+\frac{\mathcal{F}_{2}}{%
k^{2}+m_{2}^{2}}+\frac{\mathcal{F}_{3}}{k^{2}+m_{3}^{2}}+\frac{\mathcal{F}%
_{4}}{k^{2}+m_{4}^{2}}\right) \left( \delta _{\mu \nu }-\frac{k_{\mu }k_{\nu
}}{k^{2}}\right)  \label{pole2} \;, 
\end{eqnarray}
where $(m_{1},m_{2},m_{3},m_{4})$ are the roots of the denominators of expressions \eqref{pc} and \eqref{pv}.   The explicit expression of  $(m_{1},m_{2},m_{3},m_{4})$  turns out to be rather complicated, although they can be computed in closed form in terms of $(\tilde{m},\tilde{g})$. The factors $(\mathcal{R}_{1},..,\mathcal{R}_{4})$ are
\begin{align}
\mathcal{R}_{1}& =\frac{m_{1}^{4}}{%
(m_{2}^{2}-m_{1}^{2})(m_{3}^{2}-m_{1}^{2})(m_{4}^{2}-m_{1}^{2})}\;,
\label{R1} \\
\mathcal{R}_{2}& =-\frac{m_{2}^{4}}{%
(m_{2}^{2}-m_{1}^{2})(m_{3}^{2}-m_{2}^{2})(m_{4}^{2}-m_{2}^{2})}\;,
\label{R2} \\
\mathcal{R}_{3}& =\frac{m_{3}^{4}}{%
(m_{1}^{2}-m_{3}^{2})(m_{2}^{2}-m_{3}^{2})(m_{4}^{2}-m_{3}^{2})}\;,
\label{R3} \\
\mathcal{R}_{4}& =-\frac{m_{4}^{4}}{%
(m_{4}^{2}-m_{1}^{2})(m_{4}^{2}-m_{3}^{2})(m_{4}^{2}-m_{2}^{2})}  \label{R4}   \;, 
\end{align}
while, for $(\mathcal{F}_{1},..,\mathcal{F}_{4})$, we get 
\begin{align}
{\mathcal{F}}_{1}& =\frac{m_{1}^{2}\left( \mathit{\tilde{\gamma}}%
^{4}+m_{1}^{2}\left( \tilde{g}^{2}\tilde{\nu}^{2}+m_{1}^{2}\right) \right) }{%
(m_{2}^{2}-m_{1}^{2})(m_{3}^{2}-m_{1}^{2})(m_{4}^{2}-m_{1}^{2})}\;,
\label{r1} \\
{\mathcal{F}}_{2}& =-\frac{m_{2}^{2}\left( \mathit{\tilde{\gamma}}%
^{4}+m_{2}^{2}(\tilde{g}^{2}\tilde{\nu}^{2}+m_{2}^{2})\right) }{%
(m_{2}^{2}-m_{1}^{2})(m_{3}^{2}-m_{2}^{2})(m_{4}^{2}-m_{2}^{2})}\;,
\label{r2} \\
{\mathcal{F}}_{3}& =\frac{m_{3}^{4}\left( \mathit{\tilde{\gamma}}%
^{4}+m_{3}^{2}(\tilde{g}^{2}\tilde{\nu}^{2}+m_{3}^{2})\right) }{%
(m_{1}^{2}-m_{3}^{2})(m_{2}^{2}-m_{3}^{2})(m_{4}^{2}-m_{3}^{2})}\;,
\label{r3} \\
{\mathcal{F}}_{4}& =-\frac{m_{4}^{2}\left( \mathit{\tilde{\gamma}}%
^{4}+m_{4}^{2}(\tilde{g}^{2}\tilde{\nu}^{2}+m_{4}^{2})\right) }{%
(m_{1}^{2}-m_{4}^{2})(m_{3}^{2}-m_{4}^{2})(m_{4}^{2}-m_{2}^{2})}\;.
\label{r4}
\end{align}

\subsection{Analytic structure of the gauge propagator and the regimes of the theory}

In order to study the analytic structure of the gauge propagator, we look at the discriminant of the roots in the  denominator of expressions \eqref{pc},\eqref{pv}, {\it i.e.}
\begin{equation}
P(k) = \bar{k}^3+2\bar{g}  \bar{\nu} \bar{k} \left(\gamma ^4+\bar{k}^2\right)+\left(\gamma ^4+\bar{k}^2\right)^2
\end{equation} 
where we have performed the change of variables $\bar{x}=x^2$, for $x=k, \tilde{\nu},\tilde{g}$. With these new variables, the Gribov parameter reads $$\gamma^{4}=\frac{1}{4}\left( \frac{\bar{g}\bar{\nu}}{4}-\frac{\bar{g}^{2}}{9\pi ^{2}}\right).$$ Since the polynomial $P(k)$ is a quartic function of $\bar{k}$, the discriminant can be determined in a closed form, being given by
\begin{equation}
\Delta = \frac{\bar{g}^8 \left(4 {g}+9 \pi ^2 \bar{\nu}\right)^6}{2^{24} 3^{20}\pi ^{20}}  \Delta^*
\end{equation} 
with 
\begin{equation}
\begin{split}
\Delta^*= (1024 \bar{g}^6 \left(324 \pi ^4 \bar{\nu}^4+1\right)+9216 \bar{g}^5
   \left(162 \pi ^6 \bar{\nu}^5+36 \pi ^4 \bar{\nu}^3+\pi ^2 \bar{\nu}\right)- \\ 10368 \pi ^4 \bar{g}^4 \bar{\nu}^2 \left(18 \pi
   ^2 \left(567 \pi ^2 \bar{\nu}^2-8\right) \bar{\nu}^2+25\right)-46656 \pi ^4 \bar{g}^3 v \left(3420 \pi ^4
   \bar{\nu}^4+27 \pi ^2 \bar{\nu}^2+4\right)-\\8748 \pi ^4 \bar{g}^2 \left(9381 \pi ^4 \bar{\nu}^4+96 \pi ^2
   \bar{\nu}^2+4\right)-52488 \pi ^6 \bar{g}  \bar{\nu}\left(274 \pi ^2 \bar{\nu}^2+3\right)-177147 \pi ^8 \bar{\nu}^2 )
\end{split}
\end{equation} 
As it is well known, for a quartic polynomial, if $\Delta>0$,  it displays four complex roots, on the other hand if $\Delta<0$ 
the polynomial exhibits two real and two conjugate complex roots. In this way, we shall be able to characterize 
a confining and a de-confining region in the parameter space. To do that, firstly, let us note that 

\begin{equation}
\lim_{\bar{\nu}\rightarrow\infty}\Delta^* = -\infty
\end{equation}

\begin{equation}
\lim_{\bar{g}\rightarrow\infty}\Delta^* = +\infty
\end{equation}
 This means that the discriminant changes sign for either large values of the $\it{vev}$ $\nu$ or of the coupling constant $g$. The transition line of the discriminant can be plotted exactly,  as it is shown in Fig.1. In particular, in the limit when both $\bar{\nu}$ and $\bar{g}$ tend to infinity, the transition line can be approximated by taking the leading order term of the polynomial, which behaves as 

\begin{equation}
\lim_{\bar{g}\rightarrow\infty,\bar{\nu}\rightarrow\infty }\Delta^* \propto \pi^4 \bar{g}^4  \bar{\nu}^4 (4 \bar{g} - 63 \pi^2 \bar{\nu}) (4 \bar{g} + 81 \pi^2 \bar{\nu})
\end{equation}

\noindent Therefore, in the infinite limit  $\nu,g \rightarrow \infty$, with  $\frac{\bar{\nu}}{\bar{g}} =\frac{4}{63 \pi^2}$,  there is a change of sign in the discriminant, as we see from the Fig.1. \\\\Now, if we consider the limit $\bar{\nu}\rightarrow0$, we get

\begin{figure}
  \centering
    \includegraphics[width=100mm]{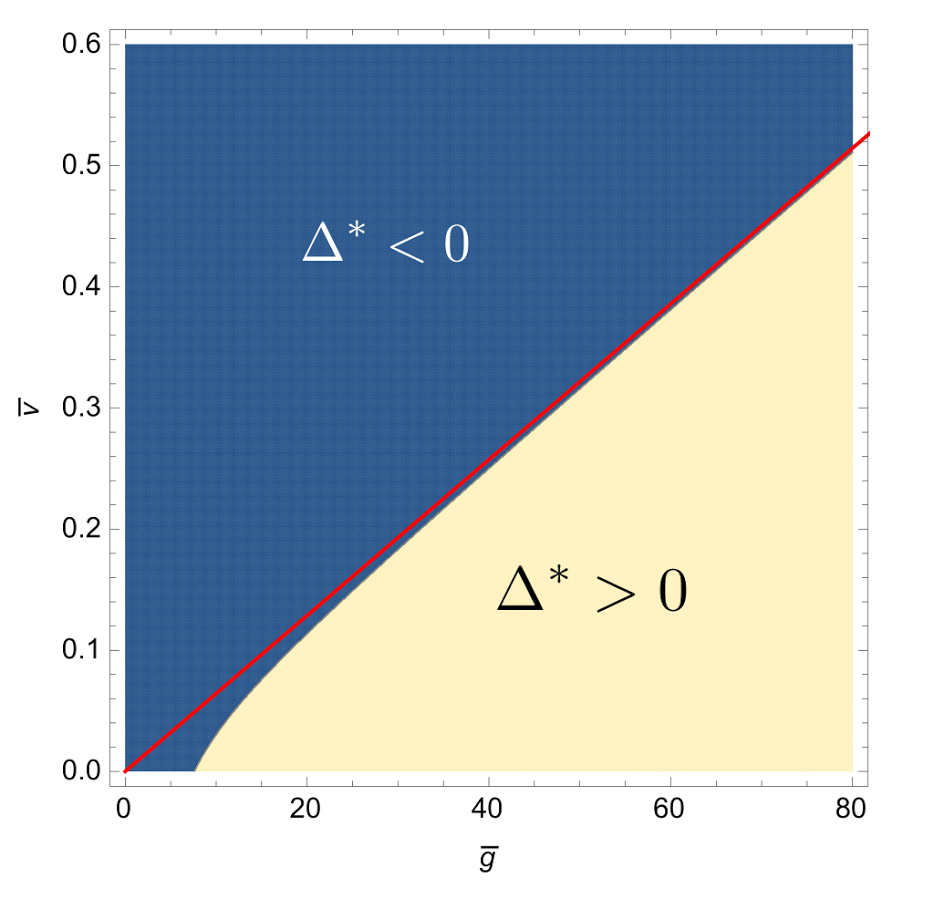}
  \caption{Threshold line of the discriminant as a function of $\bar{g}$ and $\bar{\nu}$.
  The red line represents the transition line in the limit where $\bar{g}$ and $\bar{\nu}$ tend to infinity}  \label{fig:ejemplo}
\end{figure}

\begin{figure}
  \centering
    \includegraphics[width=120mm]{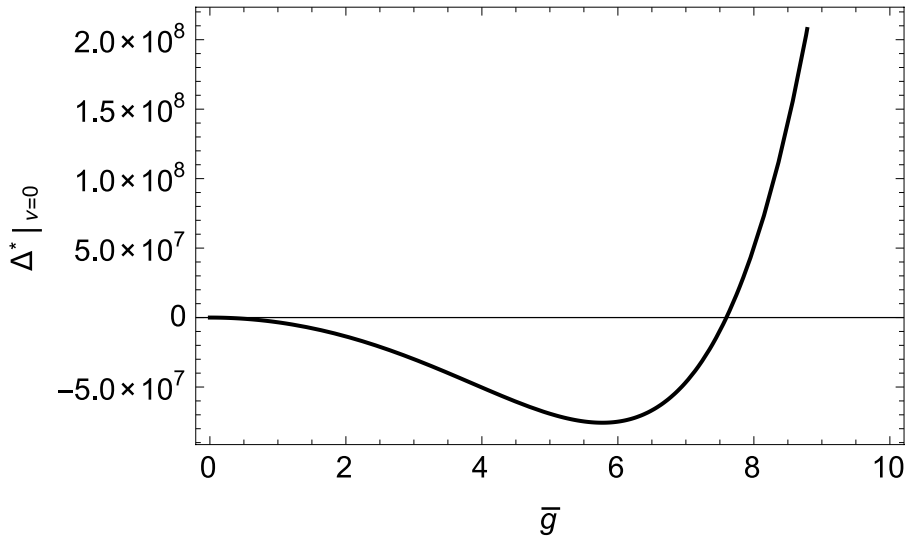}
  \caption{ $\Delta^*$ in the limit where $\bar{\nu}$ $\rightarrow$ $0$} 
  \label{fig1}
\end{figure}

\begin{equation}
\lim_{\bar{\nu}\rightarrow0}\Delta^* = 16 (64 g^6 - 2187 g^2 \pi^4)
\end{equation}

\noindent   In this case, the behaviour of the discriminant shows that the transition line starts at $$\bar{g}^*=\frac{3\ 3^{3/4} \pi }{2 \sqrt{2}}$$  

\noindent Thus, as it is apparent from Fig.1, for large values of the parameter $\bar{\nu}$ and small enough values of the parameter $\bar{g}$, we can identify two real poles, {\it e.g.} $(m_{1},m_{2})$, and two complex conjugate poles, {\it e.g.} $(m_{3},m_{4})$. This region would correspond to what we would call weak coupling region, {\it i.e.} small coupling constant and large values of the Higgs {\it vev}.  The real poles $(m_{1},m_{2})$ would correspond to Yukawa like propagators, thus being identifiable with physical excitations, provided the corresponding residues are positive. Using simple computer algebra, is easy to show that $\mathcal{R}_1$ and $\mathcal{F}_2$ are always positive, while $\mathcal{R}_2$ and $\mathcal{F}_1$ attain negative values. \\\\\
On the other hand, for large values of the coupling constant $\bar{g}$, {\it i.e.} the strong coupling regime, the propagator shows only complex poles, giving rise to the confining sector of the theory.


\section{Conclusion}

In this work we have studied the non-perturbative behaviour of the Chen-Simons-Yang-Mills in presence  of a Higgs field in the fundamental representation of the gauge group, by taking into account the Gribov horizon.   As is well known, in this representation the Higgs mechanism  affects all the components of the gauge field, giving rise to three massive gauge fields. By analysing the structure of the gauge field propagator, we have been  
able to describe confining and de-confined regions in the parameter space. More precisely, we have found that, for large values of the parameter $\bar{g}^2=\frac{g^2}{M},$ the system shows a confined regime characterized by complex poles in the gauge propagators. On the other hand, for small values of $\bar{g}$ and large enough values of the vacuum expectation value of the  Higgs filed, { $\bar{\nu}^2=\frac{\nu^2}{M}$,  it is possible to observe physical poles with positive masses in the propagator, signalling that we are in the de-confined regime of the theory. \\\\ Finally, let us also point out that the case in which the Higgs field is in the adjoint representation of the gauge group can be analysed in a similar way showing, again,  the existence of confined and de-confined regimes for the right range of parameters.


\section*{Acknowledgments}
The Conselho Nacional de Desenvolvimento Cient\'{\i}fico e
Tecnol\'{o}gico (CNPq-Brazil), the Faperj, Funda{\c{c}}{\~{a}}o de
Amparo {\`{a}} Pesquisa do Estado do Rio de Janeiro, the SR2-UERJ,  the
Coordena{\c{c}}{\~{a}}o de Aperfei{\c{c}}oamento de Pessoal de
N{\'{\i}}vel Superior (CAPES)  are gratefully acknowledged. 
\\The work of A.G is supported by FONDECYT grant Nº 3130679.

\end{document}